\documentclass[prd,twocolumn,showpacs,amsmath,amssymb]{revtex4}

\usepackage{amssymb}
\usepackage{mathrsfs}
\usepackage{txfonts}

\usepackage{graphicx}
\usepackage{dcolumn}
\usepackage{bm}

\begin{document}

\title{Non-Gaussianity in a general kind noncanonical warm inflation}
\author{Xiao-Min Zhang}
\thanks{Corresponding author}
\email{zhangxm@mail.bnu.edu.cn}
\affiliation{School of Science, Qingdao University of Technology, Qingdao 266033, China}
\author{Kai Li}
\affiliation{School of Science, Qingdao University of Technology, Qingdao 266033, China}
\author{Hong-Yang Ma}
\affiliation{School of Science, Qingdao University of Technology, Qingdao 266033, China}
\author{Qian Liu}
\affiliation{School of Science, Qingdao University of Technology, Qingdao 266033, China}
\author{Jian-Yang Zhu}
\email{zhujy@bnu.edu.cn}
\affiliation{Department of Physics, Beijing Normal University, Beijing 100875, China}
\date{\today}

\begin{abstract}
The general kind noncanonical warm inflation is introduced and its total non-Gaussianity of perturbation on uniform energy density hypersurfaces is studied. The total non-Gaussianity contains two complementary parts: the three- and the four-point correlations. The three-point correlation part denoted by $f^{int}_{NL}$, comes from the three-point correlation of inflaton field, while the four-point correlation part, denoted by $f_{NL}^{\delta N}$, is the contribution due to the four-point correlation function of the inflaton field. The aforementioned two parts of non-Gaussianity in general noncanonical warm inflation are calculated and analysed respectively, and the comparisons and discussions of the two parts are finally carried out in this study.
\end{abstract}
\maketitle

\section{\label{sec1}Introduction}
Inflation can successfully solve the problems of horizon, flatness and monopole suffered in the Big Bang Universe \cite{Guth1981,Linde1982,Albrecht1982}. Inflation is also able to give a natural mechanism to clarify the observed anisotropy of the cosmological microwave background (CMB) and the large scale structure (LSS) exactly. Till now, there are two kinds of inflationary theory: standard inflation, (also called cold inflation), and warm inflation. Warm inflation was firstly proposed by A. Berera in 1995 \cite{BereraFang,Berera1995}. Warm inflation not only inherits the advantages of standard inflation, such as solving the problems of horizon, flatness and monopole, but also has its own advantages. In warm inflation, the slow roll conditions are more easily to be satisfied due to the thermal damped effect \cite{Ian2008,Campo2010,Zhang2014,ZhangZhu}. The ``$\eta$ problem'' \cite{etaproblem,etaproblem1} and overlarge amplitude of inflaton \cite{Berera2006,BereraIanRamos} suffered in standard inflation can be cured in warm inflation scenario. The important difference between two kinds of inflationary theory is the origin of cosmological density fluctuations. The fluctuations mainly originate from thermal fluctuations in warm inflation \cite{Berera2000,Lisa2004,Taylor2000,Chris2009,BereraIanRamos}, while these fluctuations come from vacuum quantum fluctuations in standard inflation \cite{LiddleLyth,Bassett2006}. During warm inflation, the inflaton field is not isolated and the radiation can be produced constantly through the interactions between the inflaton field and other subdominated boson or fermion fields. Hence, there is no reheating phase and the Universe can smoothly go into the big-bang radiation dominated phase.

Typically, the power spectrum of density perturbations and the relative amplitude of gravitational waves are usually calculated by researchers. Because the power spectrum only reflects only two-point correlation statistics information, the information in the power spectrum is limited to some extent. A range of different inflationary models are meeting two-point correlation observations. Only calculations and observations about two-point perturbation quantities will not allow us efficiently discriminate among these models. Thus there has been growing interest in calculating not only the power spectrum but also the bispectrum and other measures of possible deviations from purely Gaussian distribution as a possible discriminant between different inflationary models. Non-Gaussianity is an important issue when analyzing inflation models. The primordial curvature perturbation in slow roll inflation is dominated by the Gaussian term, and it has a rather small deviation from pure Gaussian distributed term. The three-point function, (or its Fourier transform - the bispectrum) represents the lowest order statistics able to distinguish non-Gaussian from Gaussian perturbations \cite{Heavens1998,Ferreira1998}. The non-Gaussianity of standard inflation in both canonical inflationary theory \cite{Bartolo2004,Gangui1994,Calzetta1995,DavidLyth2005,Lyth2005,Zaballa2005} and noncanonical theory \cite{Creminelli2003,Silverstein2004,Alishahiha2004} has been analysed in previous works. It's found that the single canonical slow roll inflationary model produces a negligible non-Gaussianity, while the non-Gaussianity in noncanonical or multi-field inflationary models is significant. In warm inflation, the issue of non-Gaussianity has also been analysed in different cases, such as strong or weak dissipative regime, temperature independent or dependent case \cite{Moss2007,Gupta2002,Gupta2006,Gil2014}. Many works studied the primordial non-Gaussianity generated in warm inflationary models in recent years, but most of them have been concentrating on the canonical warm inflationary case.

The non-Gaussianity in noncanonical warm inflation was first considered in the work \cite{Zhang2015} and it reached the conclusion that small sound speed and large dissipation strength can both enhance the magnitude of the non-Gaussianity. The studied case \cite{Zhang2015} is the noncanonical warm inflation with a Lagrangian density able to be written in a separable form for the kinetic term and the potential term. In this paper, the non-Gaussianity in a more general noncanonical warm inflation is calculated where the Lagrangian density can have a coupled form for the kinetic and potential terms. Observational limits are usually put on the nonlinear parameter $f_{NL}$ and it can describe the level of non-Gaussianity effectively. The non-Gaussianity in our new noncanonical warm inflation contains two complementary parts. One part is the three-point correlation part or intrinsic part, denoted by $f^{int}_{NL}$, coming from the three-point correlation or self-interaction of the inflaton field. The other part is the four-point correlation part, denoted by $f_{NL}^{\delta N}$, which is the contribution due to the four-point correlation function of the inflaton field. The $f_{NL}^{\delta N}$ part can be calculated by $\delta N$ formalism, and $\delta N$ formalism was proposed to deal with the issue of non-Gaussianity in \cite{DavidLyth2005,Zaballa2005}. It is convenient to calculate the non-Gaussianity using $\delta N$ formalism, especially in multi-field inflationary scenario \cite{Lyth2005,Zaballa2005,Vernizzi2006,Battefeld2007,Tower2010}. We introduce $\delta N$ formalism to calculate the primordial non-Gaussianity generated in canonical warm inflation for the first time in previous study \cite{Zhang2016}, and we will use it to calculate the $\delta N$ part non-Gaussianity in noncanonical warm inflation in this work.

The paper is organized as follows: Sec. \ref{sec2} introduces a new noncanonical warm inflationary scenario, giving the basic equations of the new picture. In Sec. \ref{sec3}, we introduce the $\delta N$ formalism and calculate the part of non-Gaussianity coming from the four-point correlation of the
inflaton field, denoted by $f_{NL}^{\delta N}$. Then the three-point correlation part non-Gaussianity, denoted by $f_{NL}^{int}$, is also calculated. Finally, we draw the conclusions and discussions in Sec. \ref{sec4}.

\section{\label{sec2}the general noncanonical warm inflation}
In warm inflation, a significant amount of radiation is produced constantly during the inflationary epoch. Thus, the Universe is hot with a non-zero temperature $T$.
In warm inflation, the Universe is a multi-component system with the total matter action:
\begin{equation}\label{action}
  S=\int d^4x \sqrt{-g}  \left[\mathcal{L}(X',\varphi)+\mathcal{L}_R+\mathcal{L}_{int}\right],
\end{equation}
where $X'=\frac12g^{\mu\nu}\partial_{\mu}\varphi\partial_{\nu}\varphi$.
In above equation, $\mathcal{L}(X',\varphi)$ is the Lagrangian density of the noncanonical field, $\mathcal{L}_R$ is the Lagrangian density of radiation fields and $\mathcal{L}_{int}$ represents the interactions between the scalar fields in warm inflation. A proper noncanonical Lagrangian density should meet the following conditions: $\mathcal{L}_{X'}\geq0$ and $\mathcal{L}_{X'X'}\geq0$ \cite{Franche2010,Bean2008}, where the subscript $X'$ denotes a derivative.

Varying the action, we can get motion equation of inflaton:
\begin{equation}\label{vary}
  \frac{\partial(\mathcal{L}(X',\varphi)+\mathcal{L}_{int})}{\partial\varphi}-\left(\frac{1}{\sqrt{g}}\right)
  \partial_{\mu}\left(\sqrt{g}\frac{\partial\mathcal{L}(X',\varphi)}{\partial(\partial_{\mu}\varphi)}\right)=0.
\end{equation}
The field is homogeneous in the flat Friedmann-Robertson-Walker (FRW) Universe, i.e. $\varphi=\varphi(t)$. The motion equation of the inflaton field can be reduced to:
\begin{eqnarray}\label{EOM1}
  \left[\left(\frac{\partial\mathcal{L}(X',\varphi)}{\partial X'}+2X'\left(\frac{\partial^2\mathcal{L}(X',\varphi)}{\partial X'^2}\right)\right)\right]\ddot\varphi\nonumber\\+\left[3H\left(\frac{\partial\mathcal{L}(X',\varphi)}{\partial X'}\right)+\dot\varphi\left(\frac{\partial^2\mathcal{L}(X',\varphi)}{\partial X'\partial\varphi}\right)\right]\dot\varphi\nonumber\\-\frac{\partial(\mathcal{L}(X',\varphi)+
  \mathcal{L}_{int})}{\partial\varphi}=0,~~~~~~~~~~~~~~~~~~
\end{eqnarray}
where $X'=\frac12\dot\varphi^2$ and $H$ is the Hubble parameter.

The energy density and the pressure of inflaton are: $\rho(\varphi,X')=2X'\frac{\partial\mathcal{L}}{\partial X'}-\mathcal{L}$, $p(\varphi,X')=\mathcal{L}$. The sound speed can describe the traveling speed of scalar perturbations, defined as $c_s^2=\frac{\partial p/\partial X'}{\partial\rho/\partial X'}=\left(1+2X'\frac{\mathcal{L}_{X'X'}}{\mathcal{L}_X'}\right)^{-1}$.

Considering warm inflationary assumptions and the definition of sound speed, the motion equation of the noncanonical inflaton can be expressed as \cite{Zhang2018}:
\begin{equation}\label{EOM2}
  \mathcal{L}_{X'}c_{s}^{-2}\ddot{\varphi}+(3H\mathcal{L}_{X'}+\tilde{\Gamma})\dot{\varphi}+
  \mathcal{L}_{X'\varphi}\dot\varphi^2+V_{eff,\varphi}(\varphi,T)=0.
\end{equation}

The subscript $\varphi$ also denotes a derivative, and we write the thermal effective potential $V_{eff}$ as $V$ for simplicity. $\tilde{\Gamma}$ is the dissipative coefficient in warm inflation.

The Eq. (\ref{EOM2}) has an annoying quadratic term $\mathcal{L}_{X'\varphi}\dot\varphi^2$ due to the non-vanishing kinetic potential coupling term of Lagrangian density $\mathcal{L}_{X'\varphi}$. This makes the inflation difficult to solve such as giving slow roll approximations and calculating perturbations. Fortunately, we can eliminate this term by making a field redefinition $\phi=f(\varphi)$ in many cases \cite{Zhang2018}. In the new field representation, the Lagrangian density becomes $\mathcal{L}(X,\phi)$, where $X=\frac12\dot\phi^2$ in the FRW Universe. With the field redefinition, a new field representation $\phi=f(\varphi)$ is selected to make the coupling term disappeared:
\begin{equation}\label{coupling}
  \mathcal{L}_{X\phi}=\frac{1}{f^4_{\varphi}}[f_{\varphi}\mathcal{L}_{X'\varphi}-
  2f_{\varphi\varphi}\mathcal{L}_X'-2f_{\varphi\varphi}\mathcal{L}_{X'X'}X']=0.
\end{equation}
In this equation, $f_{\varphi}$ denotes the first-order derivative of the function $f(\varphi)$, and $f_{\varphi\varphi}$ denotes the second-order derivative of $f(\varphi)$. This procedure can be done in many general noncanonical warm inflationary cases \cite{Zhang2018}. In the redefined uncoupling $\phi$ representation, the slow roll approximations and slow roll conditions of noncanonical warm inflation can be easily given out.
The motion equation of inflaton in the new redefined $\phi$ representation becomes:
\begin{equation}\label{EOM4}
   \mathcal{L}_{X}c_{s}^{-2}\ddot{\phi}+(3H\mathcal{L}_{X}+\Gamma)\dot{\phi}+V_{eff,\phi}(\phi,T)=0,
\end{equation}
where $\Gamma$ is the dissipative coefficient in the $\phi$ representation. $\Gamma$ and $\tilde{\Gamma}$ have the relation $\tilde{\Gamma}=f_{\varphi}^2\Gamma$ \cite{Zhang2018}. With the help of field redefinition, the calculations about two-point perturbations in noncanonical warm inflation is easy to be performed \cite{Zhang2018}. Since the total non-Gaussianity is independent on the concrete field representation, the easy-to-use $\phi$ representation is chosen when calculating non-Gaussianity generated by general noncanonical warm inflation in the following.

Different from canonical warm inflation, the inflatons and the dissipative coefficient may not have the normal mass dimension in general noncanonical warm inflation.
The dissipation strength parameter $r$ in our general noncanonical warm inflation is defined as \cite{Zhang2018}:
\begin{equation}\label{r}
  r=\frac{\Gamma}{3H\mathcal{L}_X}.
\end{equation}
When $r\gg1$, warm inflation is in strong regime, while when $r\ll1$, warm inflation is in weak regime.

The slow roll parameters in inflation are usually defined as:
\begin{equation}
\tilde{\epsilon}=\frac{M_p^2}{2}\left(\frac{V_{\phi}}{V}\right) ^2, \tilde{\eta}=M_p^2\frac {V_{\phi \phi}}{V},
\tilde{\beta}=M_p^2\frac{V_{\phi}\Gamma_{\phi}}{V\Gamma}.
\end{equation}
Generally speaking, the above slow roll parameters are no longer dimensionless in general noncanonical warm inflation, different from canonical inflation. In general warm inflation, we can define some new slow roll parameters as:
\begin{equation}
\epsilon=\frac{M_p^2}{2\mathcal{L}_X}\left(\frac{V_{\phi}}{V}\right) ^2,~~~\eta=\frac{M_p^2}{\mathcal{L}_X}
\frac {V_{\phi \phi}}{V},~~~\beta=\frac{M_p^2}{\mathcal{L}_X}\frac{V_{\phi}\Gamma_{\phi}}{V\Gamma}.
\end{equation}
Now, the new slow roll parameters are dimensionless.
Associated with temperature in warm inflation, another two slow roll parameters are:
\begin{equation}
b=\frac {TV_{\phi T}}{V_{\phi}},
\end{equation}
and
\begin{equation}
c=\frac{T\Gamma_T}{\Gamma}.
\end{equation}

The validity of slow roll approximations are described by the slow roll conditions \cite{Zhang2014,Zhang2018}:
\begin{eqnarray}\label{SRcondition}
\tilde{\epsilon}\ll\frac{\mathcal{L}_{X}(1+r)}{c^2_s},~\tilde{\beta}\ll\frac{\mathcal{L}_{X}(1+r)}{c^2_s},~
\tilde{\eta}\ll\frac{\mathcal{L}_{X}}{c^2_s},\nonumber\\ b\ll\frac{min\{1,r\}}{(1+r)c^2_s},~~~~|c|<4.~~~~~~~~~~~~
\end{eqnarray}
Under slow roll conditions and in the $\phi$ representation, the motion equation of noncanonical inflaton can be reduced to:
\begin{equation}\label{SREOM}
  3H\mathcal{L}_{X}(1+r)\dot{\phi}+
  \mathcal{L}_{X\phi}\dot\phi^2+V_{eff,\phi}=0.
\end{equation}

Compared to cold inflation and canonical warm inflation, the slow roll approximations are easily to be guaranteed under the condition of noncanonical warm inflation.
The number of e-folds in general noncanonical warm inflation is expressed as:
\begin{equation}\label{efold}
 N=\int H dt=\int\frac{H}{\dot{\phi}}d\phi\simeq-\frac{1}{M_p^2}\int_{\phi_{\ast}}
^{\phi_{end}}\frac{V\mathcal{L}_X(1+r)}{V_{\phi}}d\phi,
\end{equation}
where $M_p^2=\frac 1{8\pi G}$ and the subscript $\ast$ is used to denote the time of horizon crossing.

\section{\label{sec3}the non-Gaussianity in general noncanonical warm inflation}
The non-Gaussianity in general noncanonical warm inflation contains two part: the $\delta N$ part and the intrinsic part. Now, these two part are calculated respectively.
\subsection{\label{sec31}the $\delta N$ part non-Gaussianity}

$\delta N$ formalism, often used to calculate the non-Gaussianity of multi-field inflation \cite{Vernizzi2006,Battefeld2007,Tower2010}, was proposed in many works such as \cite{Lyth2005,DavidLyth2005,Starobinsky,Sasaki1996,Sasaki1998}. The primordial curvature perturbation on uniform density hypersurfaces of the Universe, denoted by $\zeta$, was Gaussian term dominated with a nearly scale-invariant spectrum, as cosmological observations suggests.

During inflationary period, the evolution of the Universe was supposed to be determined mainly by one or more inflaton fields. Considering small perturbations, we can expand full scalar field as $\Phi_i(t,\mathbf{x})=\phi_i(t)+\delta\phi_i(t,\mathbf{x})$ in a convenient flat slicing gauge. As the observations suggests, the curvature perturbation $\zeta$ is almost Gaussian. So we expand $\zeta$ up to second-order for a good accuracy:
\begin{equation}\label{zeta2}
  \zeta(t,\mathbf{x})=\delta N\simeq \sum_i N_{,i}(t)\delta\phi_i+\frac12\sum_{ij} N_{,ij}(t)\delta\phi_i\delta\phi_j,
\end{equation}
where $N_{,i}\equiv\frac{\partial N}{\partial\phi_i}$ and $N_{,ij}\equiv\frac{\partial^2 N}{\partial\phi_i\partial\phi_j}$.

They may be entirely responsible for the observed non-Gaussianity if the field perturbations are pure Gaussian. This non-Gaussianity reflects the contributions of four-point correlations, called $\delta N$ part non-Gaussianity.

However, the inflaton field perturbation in inflation deviates from pure Gaussian distribution to some extent. The deviation from pure Gaussian distribution in noncanonical inflation is larger than the one in canonical inflation. Thus in general noncanonical warm inflation, we also need to compute non-Gaussianity generated by intrinsic non-Gaussianity of inflaton field.

The power spectrum of the curvature perturbation $\zeta$, denoted by $\mathcal{P}_{\zeta}$, is defined in the relation:
\begin{equation}\label{spectrum}
  \langle\zeta_{\mathbf{k}_1}\zeta_{\mathbf{k}_2}\rangle\equiv(2\pi)^3\delta^3(\mathbf{k}_1+\mathbf{k}_2)
\frac{2\pi^2}{k_1^3}\mathcal{P}_{\zeta}(k_1),
\end{equation}
and $\mathcal{P}_{\zeta}(k)\equiv\frac{k^3}{2\pi^2}P_{\zeta}(k)$.

Three-point function of curvature perturbation, or its Fourier transform, the bispectrum, is the dominated and lowest order non-Gaussianity. The bispectrum is usually defined through the relation:
\begin{equation}\label{bispectrum0}
  \langle\zeta_{\mathbf{k}_1}\zeta_{\mathbf{k}_2}\zeta_{\mathbf{k}_3}\rangle\equiv(2\pi)^3\delta^3
(\mathbf{k}_1+\mathbf{k}_2+\mathbf{k}_3)B_{\zeta}(k_1,k_2,k_3).
\end{equation}

Observational limits are usually put on the nonlinear parameter $f_{NL}$. And the magnitude of non-Gaussianity can be evaluated through the nonlinear parameter. Since the curvature perturbations are Gaussian term dominated, its bispectrum can have the general form:
\begin{equation}\label{Bzeta}
  B_{\zeta}(k_1,k_2,k_3)\equiv -\frac65f_{NL}(k_1,k_2,k_3)\left[P_{\zeta}(k_1)P_{\zeta}(k_2)+cyclic \right].
\end{equation}

We can use the ralation Eq. (\ref{zeta2}) to compute the whole non-Gaussianity, and the calculation can yield
\begin{eqnarray}\label{threepoint}
  \langle\zeta_{\mathbf{k}_1}\zeta_{\mathbf{k}_2}\zeta_{\mathbf{k}_3}\rangle =\sum_{ijk}N_{,i}N_{,j}N_{,k}
  \langle\delta\phi^{i}_{\mathbf{k}1}\delta\phi^{j}_{\mathbf{k}2}\delta\phi^{k}_{\mathbf{k}3}\rangle ~~~~~~~~~~~\nonumber\\
  +\frac12\sum_{ijkl}N_{,i}N_{,j}N_{,kl}\langle\delta\phi^{i}_{\mathbf{k}1}\delta\phi^{j}_{\mathbf{k}2}
  (\delta\phi^{k}\star\delta\phi^{l})_{\mathbf{k}3}\rangle+perms.
\end{eqnarray}
In above equation, a star denotes the convolution, and the correlation functions higher than four-point are neglected. The first line in above equation, a three-point correlation, is the contribution from the intrinsic non-Gaussianity of the inflaton fields. The second line, a four-point correlation, reflects the $\delta N$ part non-Gaussianity, which is scale independent. It can be calculated conveniently by using $\delta N$ formalism.

According to the $\delta N$ formalism, the expression for $\delta N$ part nonlinear parameter is given by \cite{Lyth2005,Boubekeur}:
\begin{equation}\label{fNL}
  -\frac35 f_{NL}^{\delta N}=\frac{\sum_{ij}N_{,i}N_{,j}N_{,ij}}{2\left[\sum_i N^2_{,i}\right]^2}.
\end{equation}

During noncanonical warm inflation, only one inflaton field dominates, so only one $\delta\phi_i$ is relevant. Then Eq. (\ref{zeta2}) reduces to
\begin{equation}\label{zeta3}
  \zeta(t,\mathbf{x})=N_{,i}\delta\phi_i+\frac12 N_{,ii}\left(\delta\phi_i\right)^2,
\end{equation}
so we can yield
\begin{equation}\label{fNL1}
  -\frac35f_{NL}^{\delta N}=\frac12\frac{N_{,ii}}{N^2_{,i}}.
\end{equation}
Since there is only one $\delta\phi_i$, without ambiguity, we can rewrite $N_{,i}$ as $N_{\phi}$ and $N_{,ii}$ as $N_{\phi\phi}$ below.
The $f_{NL}^{\delta N}$ term is scale independent can be found through Eq. (\ref{fNL1}). The total non-Gaussianity should be described by $f_{NL}=f_{NL}^{\delta N}+f^{int}_{NL}$, and our calculation is based on the relation.

Inflationary observations are estimated on the time of horizon crossing. Since horizon crossing is well inside the slow roll inflationary regime, it is safe to calculate the $\delta N$ part nonlinear parameter $f_{NL}^{\delta N}$ in slow roll approximations. Under slow roll conditions, this relation can be given out:
\begin{equation}\label{Nphi}
  N_{\phi}=-\frac{1}{M_p^2}\frac{V\mathcal{L}_X(1+r)}{V_{\phi}}.
\end{equation}

From Eq. (\ref{Nphi}), we can get
\begin{eqnarray}\label{Nphiphi}
  N_{\phi\phi}=-\frac{1}{M_p^2}\left[\mathcal{L}_{X}(1+r)-\frac{VV_{\phi\phi}\mathcal{L}_{X}(1+r)}{V_{\phi}^2}
  \right.\nonumber\\ \left.+\frac{V\Gamma_{\phi}}{3HV_{\phi}}-\frac{\mathcal{L}_{X}r}{2}\right].~~~~~~~~~~~~~~~~~~~~~~
\end{eqnarray}
And then we have
\begin{eqnarray}\label{NphiNphiphi}
  \frac{N_{\phi\phi}}{N_{\phi}^2}=-M_p^2\left[\frac{V_{\phi}^2}{\mathcal{L}_X(1+r)V^2}-\frac{V_{\phi\phi}}
  {\mathcal{L}_X(1+r)V}\right.\nonumber\\ ~~~~~~~~~~~~\left.+\frac{\Gamma_{\phi}V_{\phi}r}{\Gamma V\mathcal{L}_X(1+r)^2}-\frac{rV_{\phi}^2}{2V^2\mathcal{L}_X(1+r)^2}\right].~~~~~~
\end{eqnarray}

With the help of the Eqs. (\ref{fNL1}) and (\ref{NphiNphiphi}), and the slow roll parameters $\tilde{\epsilon}=\frac{M_p^2}{2}\left(\frac{V_{\phi}}{V}\right) ^2$, $\tilde{\eta}=M_p^2\frac {V_{\phi \phi}}{V}$, $\tilde{\beta}=M_p^2\frac{V_{\phi}\Gamma_{\phi}}{V\Gamma}$, or the new defined slow roll parameters $\epsilon=\frac{M_p^2}{2\mathcal{L}_X}\left(\frac{V_{\phi}}{V}\right)^2$, $\eta=\frac{M_p^2}{\mathcal{L}_X}
\frac {V_{\phi \phi}}{V}$, $\beta=\frac{M_p^2}{\mathcal{L}_X}\frac{V_{\phi}\Gamma_{\phi}}{V\Gamma}$, we can obtain the $\delta N$ part nonlinear parameter:
\begin{eqnarray}\label{fNLdeltaN}
  f_{NL}^{\delta N}&=&\frac56\left[\frac{2\tilde{\epsilon}}{\mathcal{L}_X(1+r)}-\frac{\tilde{\eta}}{\mathcal{L}_X(1+r)}+
  \frac{r\tilde{\beta}}{\mathcal{L}_X(1+r)^2}-\frac{r\tilde{\epsilon}}{\mathcal{L}_X(1+r)^2}\right]\nonumber\\
  &=&\frac56\left[\frac{2\epsilon}{1+r}-\frac{\eta}{1+r}+
  \frac{r\beta}{(1+r)^2}-\frac{r\epsilon}{(1+r)^2}\right]\ll1.
\end{eqnarray}

From above calculations, we can reach the conclusion that the $\delta N$ part nonlinear parameter $f_{NL}^{\delta N}$ is scale independent. This is due to that the $\delta N$ part nonlinear parameter is decided only by nonperturbative background equations, as $\delta N$ formalism indicated. From the slow roll conditions in noncanonical warm inflation and above equation, we can find that $|f_{NL}^{\delta N}|\sim \mathcal{O}\left(\frac{\tilde{\epsilon}}{\mathcal{L}_{X}(1+r)}\right)\lesssim1$. Thus $f_{NL}^{\delta N}$ is a first order small quantity in slow roll approximations.

The slow roll conditions in noncanonical warm inflation suggest that the amplitude of $\delta N$ part non-Gaussianity is not significant during inflationary epoch, and it can grow slightly along with the inflation of Universe. Since the $\delta N$ form non-Gaussianity is not large enough, obviously, it's insufficient to use this part to represent the whole primordial non-Gaussianity generated by inflation as performed in some papers \cite{Zhang2016,DavidLyth2005}. It is essential to calculate the intrinsic non-Gaussianity generated by three-point correlation functions of inflaton field.

\subsection{\label{sec32}the intrinsic part non-Gaussianity}
Different to cold inflation, the fluctuations in warm inflation origin from thermal fluctuations.
In noncanonical warm inflation, only one scalar field acts as inflaton. And considering small perturbations, we can expand the full inflaton as $\Phi(\mathbf{x},t)=\phi(t)+\delta\phi(\mathbf{x},t)$, where $\delta\phi$ is the perturbation field around the homogenous background field $\phi(t)$ as usual.

Perturbation observation quantities are evaluated at the time of horizon crossing, and horizon crossing is well inside slow roll regime. In slow roll noncanonical warm inflation, due to the enhanced Hubble damped term and thermal damped term, the evolution of inflaton is overdamped. The evolution of the inflaton perturbations is quite slow in slow roll regime as indicated in \cite{Lisa2004}. So introducing the white stochastic thermal noise $\xi$, the motion equation of the full field in Fourier space can be expressed as:
\begin{equation}\label{EOMphik}
\frac{d\Phi(\mathbf{k},t)}{dt}=\frac{1}{3H\mathcal{L}_X(1+r)}\left[-k_p^2\mathcal{L}_X\delta\phi(\mathbf{k},t)-V_{\phi}
(\Phi(\mathbf{k},t))+\xi(\mathbf{k},t)\right].
\end{equation}
In above equation, $\mathbf{k}$ is the physical momentum with the notation $\mathbf{k}\equiv\mathbf{k}_p=\mathbf{k}_c/a$ ($\mathbf{k}_c$ is the comoving momentum). And the magnitude of physical momentum is denoted as $k=|\mathbf{k}|$ in the following.
Then the motion equation for the perturbations can be obtained from Eq. (\ref{EOMphik}):
\begin{equation}\label{SREOM1}
  3H\mathcal{L}_X(1+r)\delta\dot{\phi}_k+
  (\mathcal{L}_X k^2+V_{\phi\phi})\delta\phi_k=\xi_k.
\end{equation}
The thermal stochastic noise $\xi$ in thermal system has zero mean $\langle\xi\rangle=0$.
The noise source is Markovian in high temperature limit $T\rightarrow\infty$: $\langle\xi(\mathbf{k},t)\xi(\mathbf{k'},t')\rangle=2\Gamma T(2\pi)^3\delta^3(\mathbf{k}-\mathbf{k'})\delta(t-t')$ \cite{Lisa2004,Gleiser1994}.

Thermal noise term introduced in warm inflation is white noise, i.e. Gaussian distributed  \cite{Berera2000}. The inflaton perturbations can be divided into first order, second order and higher orders (higher order perturbations are tiny and thus we do not concentrate on them). The leading order inflaton perturbation is first order perturbation. It is the linear response to the thermal noise, so it is also Gaussian distributed. The inflaton fluctuations should be expanded at least to second order if we want to calculate the predicted bispectrum of inflaton perturbation from Eq. (\ref{EOMphik}). The inflaton perturbation is expanded as: $\delta\phi(\mathbf{x},t)=\delta\phi_1(\mathbf{x},t)+\delta\phi_2(\mathbf{x},t)$, where $\delta\phi_1=\mathcal{O}(\delta\phi)$ and $\delta\phi_2=\mathcal{O}(\delta\phi^2)$.

Then we can get the motion equation for the first and second order perturbations through perturbing the evolution equation of the full inflaton Eq. (\ref{EOMphik}) to second order:
\begin{widetext}
\begin{equation}\label{deltaphi1}
\frac{d}{dt}\delta\phi_1(\mathbf{k},t)=\frac{1}{3H\mathcal{L}_{X}(1+r)}\left[-\mathcal{L}_Xk^2
\delta\phi_1(\mathbf{k},t)-V_{\phi\phi}(\phi(t))\delta\phi_1(\mathbf{k},t)+\xi(\mathbf{k},t)\right],
\end{equation}
and
\begin{eqnarray}\label{deltaphi2}
\frac{d}{dt}\delta\phi_2(\mathbf{k},t)&=&\frac{1}{3H\mathcal{L}_{X}(1+r)}\left[-\mathcal{L}_Xk^2
\delta\phi_2(\mathbf{k},t)-V_{\phi\phi}
(\phi(t))\delta\phi_2(\mathbf{k},t)\right.\nonumber\\ &-&\left.\frac12V_{\phi\phi\phi}(\phi(t))\int\frac{dp^3}{(2\pi)^3}\delta\phi_1(\mathbf{p},t)\delta\phi_1(\mathbf{k}
-\mathbf{p},t)-k^2\mathcal{L}_{XX}\int\frac{dp^3}{(2\pi)^3}\delta\phi_1(\mathbf{p},t)\delta X_1(\mathbf{k}
-\mathbf{p},t)\right].
\end{eqnarray}

Solving the above two evolution equations respectively, we can get the analytic solutions of first order perturbations $\delta\phi_1$ and second order perturbations $\delta\phi_2$:
\begin{eqnarray}\label{solution1}
\delta\phi_1(\mathbf{k},\tau)&=&\frac{1}{3H\mathcal{L}_X(1+r)}\exp\left[-\frac{\mathcal{L}_Xk^2+m^2}
{3H\mathcal{L}_X(1+r)}\left(\tau-\tau_0\right)\right]
\int_{\tau_0}^{\tau}d\tau'\exp\left[\frac{\mathcal{L}_Xk^2+m^2}{3H\mathcal{L}_X(1+r)}\left(\tau'-\tau_0\right)\right]
\xi(\mathbf{k},\tau')\nonumber\\&+&
\delta\phi_1(\mathbf{k},\tau_0)\exp\left[-\frac{\mathcal{L}_Xk^2+m^2}
{3H\mathcal{L}_X(1+r)}\left(\tau-\tau_0\right)\right],
\end{eqnarray}
and
\begin{eqnarray}\label{solution2}
\delta\phi_2(\mathbf{k},\tau)&=&\exp\left[-\frac{\mathcal{L}_Xk^2+m^2}
{3H\mathcal{L}_X(1+r)}\left(\tau-\tau_0\right)\right]\int_{\tau_0}^{\tau}d\tau'\exp
\left[\frac{\mathcal{L}_Xk^2+m^2}
{3H\mathcal{L}_X(1+r)}\left(\tau'-\tau_0\right)\right]\left[A(k,\tau')
\int\frac{dp^3}{(2\pi)^3}\delta\phi_1(\mathbf{p},\tau')
\delta\phi_1(\mathbf{k}-\mathbf{p},\tau')\right.\nonumber\\ &-&\left. \frac{k^2\mathcal{L}_{XX}\sqrt{2X}}{[3H\mathcal{L}_X(1+r)]^2}\int\frac{dp^3}{(2\pi)^3}\delta\phi_1(\mathbf{p},\tau')
\xi(\mathbf{k}-\mathbf{p},\tau')\right] + \delta\phi_2(\mathbf{k},\tau_0)\exp\left[-\frac{\mathcal{L}_Xk^2+m^2}
{3H\mathcal{L}_X(1+r)}\left(\tau-\tau_0\right)\right],
\end{eqnarray}
\end{widetext}
where $m^2=V_{\phi\phi}$ is the effective squared inflaton mass and the parameter $A(k,\tau)$ appeared in above equation is
\begin{equation}\label{B}
  A(k,\tau)=-\frac{V_{\phi\phi\phi}}{6H\mathcal{L}_X(1+r)}+\frac{k^2\mathcal{L}_{XX}\sqrt{2X}(\mathcal{L}_Xk^2+
  V_{\phi\phi})}{[3H\mathcal{L}_X(1+r)]^2}.
\end{equation}

Both the second terms on the righthand side of above solutions are ¡°memory¡± terms that reflect the state of the given mode at the beginning time $\tau_0$.
From Eq. (\ref{solution1}) we can find that the larger the squared magnitude of the physical momentum $k^2$ is, the faster the relaxation rate is. If $k^2$ is large enough for the mode to relax within a Hubble time, that mode thermalizes. While as soon as the physical momentum of a $\Phi(\mathbf{x},t)$ field mode becomes less than a critical value $k_F$, it essentially feels no effect of the thermal noise during a Hubble time. $k_F$ is called freeze-out momentum in warm inflation. Based on Eq. (\ref{solution1}), we define a parameter $\tau(\phi)=\frac{3H\mathcal{L}_X(1+r)}{\mathcal{L}_Xk_p^2+m^2}$ that describe the efficiency of the thermalizing process. Then the freeze-out momentum $k_F$ can be given out by the condition:
\begin{equation}\label{kf}
   \frac{\mathcal{L}_{X}k_F^2+m^2}{(3H\mathcal{L}_{X}(1+r)H}=1.
\end{equation}
From the criterion, we can work out that $k_F\simeq\sqrt{3(1+r)}H$.

As we stated before, the first order inflaton perturbations $\delta\phi_1$ are Gaussian fields. From the stochastic statistics properties, we know their bispectrum vanishes. If we want to calculate non-Gaussianity, the leading order should be the bispectrum generated from two first order and one second order fluctuations:
\begin{widetext}
\begin{eqnarray}\label{threepoint1}
&&\langle\delta\phi(\mathbf{k}_1,\tau)\delta\phi(\mathbf{k}_2,\tau)\delta\phi(\mathbf{k}_3,\tau)\rangle \nonumber\\ &=& \left\{\int_{\tau_0}^{\tau} d\tau' \exp\left[\frac{\mathcal{L}_Xk_3^2+m^2}
{3H\mathcal{L}_X(1+r)}\left(\tau'-\tau_0\right)\right]
\left[A(k_3,\tau')\int\frac{dp^3}{(2\pi)^3}\langle\delta\phi_1(\mathbf{k}_1,\tau)\delta\phi_1(\mathbf{k_2},\tau)
\delta\phi_1(\mathbf{p},\tau')
\delta\phi_1(\mathbf{k}_3-\mathbf{p},\tau')\rangle\right.\right.\nonumber\\ &-&\left.\left. \frac{k_3^2\mathcal{L}_{XX}\sqrt{2X}}{[3H\mathcal{L}_X(1+r)]^2}\int \frac{dp^3}{(2\pi)^3}\langle\delta\phi_1(\mathbf{k}_1,\tau)\delta\phi_1(\mathbf{k_2},\tau)\delta\phi_1(\mathbf{p},\tau')
\xi(\mathbf{k}_3-\mathbf{p},\tau')\rangle\right] \right\}\times \exp\left[-\frac{\mathcal{L}_Xk_3^2+m^2}
{3H\mathcal{L}_X(1+r)}\left(\tau-\tau_0\right)\right] \nonumber\\&+&\exp\left[-\frac{\mathcal{L}_Xk_3^2+m^2}
{3H\mathcal{L}_X(1+r)}\left(\tau-\tau_0\right)\right] \langle\delta\phi_1(\mathbf{k}_1,\tau)\delta\phi_1(\mathbf{k_2},\tau)\delta\phi_2
(\mathbf{k}_3,\tau_0)\rangle+(\mathbf{k}_1\leftrightarrow\mathbf{k}_3)+(\mathbf{k}_2\leftrightarrow\mathbf{k}_3).
\end{eqnarray}
\end{widetext}

The amplitude of the bispectrum are determined at the time when cosmological scale exits the horizon, nearly 60 e-folds before the end of inflation and for $\mathbf{k}_1$, $\mathbf{k}_2$ and $\mathbf{k}_3$ all within a few e-folds of crossing the horizon.
From the expression of $k_F$, we know $k_F>H$ in noncanonical warm inflationary model. It implies
the $\delta\phi$ correlations that should be computed at the time of Hubble horizon crossing $k=H$, are the thermalized correlations that were fixed at earlier freeze-out time $k=k_F$ \cite{Berera2000}. So the time interval in the corrections can be given by
\begin{equation}\label{time}
\Delta t_F=t_H-t_F\simeq\frac{1}{H}\ln\left(\frac{k_F}{H}\right).
\end{equation}
Then the bispectrum can reduce to
\begin{eqnarray}\label{threepoint2}
&&\langle \delta \phi ({\bf k}_1,t)\delta \phi ({\bf k}_2,t)\delta \phi (%
{\bf k}_3,t)\rangle \simeq 2A(k_F,t_F)\Delta t_F  \nonumber \\
&&\times \left[ \int \frac{dp^3}{(2\pi )^3}\langle \delta \phi _1({\bf k}%
_1,t)\delta \phi _1({\bf p},t)\rangle \langle \delta \phi _1({\bf k}%
_2,t)\delta \phi _1({\bf k}_3-{\bf p},t)\rangle \right.   \nonumber \\
&&+\left. ({\bf k}_1\leftrightarrow {\bf k}_3)+({\bf k}_2\leftrightarrow
{\bf k}_3)\right].
\end{eqnarray}

In spatially flat gauge we have the relation:
\begin{equation}\label{zeta}
\zeta=\frac{H}{\dot\phi}\delta\phi.
\end{equation}

Through the Eqs. (\ref{threepoint2}) (\ref{zeta}) and (\ref{Bzeta}), we can obtain the nonlinear parameter of the intrinsic non-Gaussianity:
\begin{eqnarray}\label{fNLint}
f_{NL}^{int}=&-&\frac56\frac{\dot\phi}{H}2A(k_F,t_F)\Delta t_F \nonumber\\ =&-&\frac56
\ln\sqrt{3(1+r)}\left[\frac{\tilde{\epsilon}\tilde{\varepsilon}}{\mathcal{L}_X^2(1+r)^2} +\left(\frac{1}{c_s^2}-1\right)\right] \nonumber\\=&-&\underbrace{\frac56
\ln\sqrt{3(1+r)}\frac{\epsilon\varepsilon}{(1+r)^2}}_{term1} \nonumber\\ &-&\underbrace{\frac56
\ln\sqrt{3(1+r)}\left(\frac{1}{c_s^2}-1\right)}_{term2}.
\end{eqnarray}
The parameter $\varepsilon$ in above equation is defined as $\varepsilon=2M_p^2\frac{V_{\phi\phi\phi}}{\mathcal{L}_XV_{\phi}}$, which can be seen as a slow roll parameter that has the same magnitude as the slow roll parameter $\eta$. Hence, the intrinsic nonlinear parameter $f_{NL}^{int}$ in the context of general noncanonical warm inflationary theory is obtained, able to be compared to the observations.

Using the slow roll conditions in general noncanonical warm inflation, we can get:
\begin{equation}\label{term1}
  term 1=-\frac56\frac{\epsilon\varepsilon}{(1+r)^2}\ln\sqrt{3(1+r)}\ll1.
\end{equation}
Obviously, the first term of intrinsic nonlinear parameter $f_{NL}^{int}$ is much less than the second term, so the dominating term in $f_{NL}^{int}$ is the second term. Then the intrinsic nonlinear parameter can be approximated to
\begin{equation}\label{intrinsic}
  f_{NL}^{int}\simeq-\frac56\left(\frac{1}{c_s^2}-1\right)\ln\sqrt{3(1+r)}.
\end{equation}
Based on this equation, a small sound speed of the inflaton can much enhance the amount of the intrinsic non-Gaussianity. It means that the stronger the noncanonical effect is, the more the departure from Gaussian distribution is. From the result, we also conclude that the strong dissipation effect of warm inflation can also contribute to the intrinsic non-Gaussianity to some degree. The new observations of PLANCK satellite suggest no evidence for primordial non-Gaussianity \cite{PLANCKNG2015} and place an upper bound $|f_{NL}|<\mathcal{O}(10^2)$, which means the dominating factor of intrinsic non-Gaussianity, i.e. sound speed $c_s$ should not be too small.

\subsection{\label{sec33}the estimation of the whole non-Gaussianity}

Based on the results we reached in preceding part of the paper, now, we compare the non-Gaussianities in two parts. As we stated before, $f_{NL}^{\delta N}$ can be expressed as the polymerization of the new defined slow roll parameters. Therefore, $f_{NL}^{\delta N}$ can not be much larger than one. While the intrinsic part $f^{int}_{NL}$ is much larger than one if the sound speed of the inflaton is small enough. Then we can get the conclusion that the dominating part in the whole non-Gaussianity is the intrinsic part when noncanonical effect is strong ($c_s\ll1$). The strong noncanonical effect (i.e. small sound speed) contributes most to the non-Gaussian distributions in noncanonical warm inflation.

With the nonlinear parameter in two parts we have got, the total nonlinear parameter now can be approximated as:
\begin{equation}\label{fNLtotal}
  f_{NL}=f_{NL}^{int}+f_{NL}^{\delta N}\simeq\left(\frac{1}{c_s^2}-1\right)\ln\sqrt{3(1+r)}.
\end{equation}

The obtained general non-Gaussian results in noncanonical warm inflation can reduce to canonical case in the limit $c_s\rightarrow1$:
\begin{equation}\label{fnlncanonical}
  f_{NL}^{\delta N}=\frac{5\epsilon}{3(1+r)}-\frac{5\eta}{6(1+r)}-\frac{5r\epsilon}{6(1+r)^2}+\frac{5r\beta}{6(1+r)^2},
\end{equation}
and
\begin{equation}\label{fnlintcanonical}
  f_{NL}^{int}=-\frac56\ln\sqrt{3(1+r)}\frac{\epsilon\varepsilon}{(1+r)^2}.
\end{equation}
The term $f^{\delta N}_{NL}$ is a first order slow roll small quantity while $f_{NL}^{int}$ is a second order small quantity. Above equations suggest that in canonical warm inflation, the non-Gaussian feature is quite different from that in noncanonical warm inflation. The differences are mainly reflected by that, in canonical warm inflation, the total nonlinear parameter $f_{NL}$ is dominated by the $f_{NL}^{\delta N}$ term instead of the $f_{NL}^{int}$ term. Eqs. (\ref{fnlncanonical}) and (\ref{fnlintcanonical}) suggest that primordial non-Gaussianity in canonical warm inflation is not significant, quite different from noncanonical warm inflation.

\section{\label{sec4}conclusions and discussions}

In this paper, the whole primordial non-Gaussianity generated in general noncanonical warm inflation was investigated. We introduce the framework of general noncanonical warm inflation. The main equations of noncanonical warm inflation, such as motion equation, e-folds, slow roll equations and slow roll conditions are presented. Also, we concentrate on the major issue: non-Gaussianity generated by general warm inflation. The nonlinear parameter are usually used to measure the magnitude of non-Gaussianity and it can be divided into two parts: the intrinsic part $f_{NL}^{int}$ and the $\delta N$ part $f^{\delta N}_{NL}$. The first part describes the contribution of the three-point correlation, i.e., the intrinsic non-Gaussianity of the inflaton field. And the second part results from the four-point correlation of inflaton perturbations. These two parts together can represent the primordial non-Gaussianity in noncanonical warm inflation entirely.

To calculate the $\delta N$ part non-Gaussianity, $\delta N$ formalism is introduced and used. With the obtained result, we reach the conclusion that $f_{NL}^{\delta N}$ can be expressed as a linear combination of the new defined slow roll parameters. So $f_{NL}^{\delta N}$ is a first order small quantity in slow roll inflation, which indicates the $\delta N$ part non-Gaussianity in general noncanonical warm inflation can not be significant. This case is the same as in canonical warm inflation. But the situation is different when calculating intrinsic non-Gaussianity. The intrinsic part non-Gaussianity origins from three-points correlations of inflaton itself, mainly dependent on the sound speed and dissipation strength parameter. Throughout the whole non-Gaussianity in general noncanonical warm inflation, we find that $f_{NL}^{int}$ overwhelms the $f_{NL}^{\delta N}$ part, and the sound speed plays the dominated role. Thermal dissipation effects and higher order correlations also contribute to non-Gaussianity to a certain extent.

Warm k-inflation is no longer dominated by the potential, and its non-Gaussianity also deserves more cognitions and researches. In the near future we will also concentrate on this problem.

\section{Acknowledgments}
This work was supported by the National Natural Science Foundation of China (Grant No. 11605100, No. 11704214 and No. 11975132).

\end{document}